\documentclass[onecolumn,amsmath,showpacs,11pt]{revtex4}
\usepackage{graphicx}
\usepackage{dcolumn}
\usepackage{bm}
\begin{document}
\newcommand{\hs}{\hspace*{0.5cm}}
\newcommand{\vs}{\vspace*{0.5cm}}
\newcommand{\be}{\begin{equation}}
\newcommand{\ee}{\end{equation}}
\newcommand{\bea}{\begin{eqnarray}}
\newcommand{\eea}{\end{eqnarray}}
\newcommand{\ben}{\begin{enumerate}}
\newcommand{\een}{\end{enumerate}}
\newcommand{\bde}{\begin{widetext}}
\newcommand{\ede}{\end{widetext}}
\newcommand{\nn}{\nonumber}
\newcommand{\crn}{\nonumber \\}
\newcommand{\Tr}{\mathrm{Tr}}
\newcommand{\non}{\nonumber}
\newcommand{\noi}{\noindent}
\newcommand{\al}{\alpha}
\newcommand{\la}{\lambda}
\newcommand{\bet}{\beta}
\newcommand{\ga}{\gamma}
\newcommand{\va}{\varphi}
\newcommand{\om}{\omega}
\newcommand{\pa}{\partial}
\newcommand{\+}{\dagger}
\newcommand{\fr}{\frac}
\newcommand{\bc}{\begin{center}}
\newcommand{\ec}{\end{center}}
\newcommand{\Ga}{\Gamma}
\newcommand{\de}{\delta}
\newcommand{\De}{\Delta}
\newcommand{\ep}{\epsilon}
\newcommand{\varep}{\varepsilon}
\newcommand{\ka}{\kappa}
\newcommand{\La}{\Lambda}
\newcommand{\si}{\sigma}
\newcommand{\Si}{\Sigma}
\newcommand{\ta}{\tau}
\newcommand{\up}{\upsilon}
\newcommand{\Up}{\Upsilon}
\newcommand{\ze}{\zeta}
\newcommand{\ps}{\psi}
\newcommand{\Ps}{\Psi}
\newcommand{\ph}{\phi}
\newcommand{\vph}{\varphi}
\newcommand{\Ph}{\Phi}
\newcommand{\Om}{\Omega}

\title{Left-right model for dark matter}  

\author{P. V. Dong}\email{pvdong@iop.vast.vn}\affiliation{Institute of Physics, Vietnam Academy of Science and Technology, 10 Dao Tan, Ba Dinh, Hanoi, Vietnam}
\author{D. T. Huong}\email{dthuong@iop.vast.vn}\affiliation{Institute of Physics, Vietnam Academy of Science and Technology, 10 Dao Tan, Ba Dinh, Hanoi, Vietnam}

\begin{abstract}
We argue that dark matter can automatically arise from a gauge theory that possesses a non-minimal left-right gauge symmetry, $SU(3)_C\otimes SU(M)_L\otimes SU(N)_R\otimes U(1)_X$, for $(M,N)=(2,3),\ (3,2),\ (3,3),\cdots,$ and $(5,5)$.          
\end{abstract}

\date{\today}

\maketitle

{\it Introduction}: The astrophysical and cosmological observations have implied that our universe is made of roundly 26\% dark matter \cite{planck}. If its present-day existence is a result of thermal relics of the early universe, it may be formed of weakly-interacting massive particles (WIMPs), which do not present in the standard model \cite{pdg}. What is the nature of WIMPs? In this work, we show that the left-right models which generalize the electroweak symmetry to non-minimal left-right gauge groups can automatically provide the WIMPs from their gauge principles.         

{\it Proposal of the model}: Consider the model based on the gauge group, 
\be SU(3)_C\otimes SU(M)_L\otimes SU(N)_R\otimes U(1)_X,\ee 
which explicitly violates a symmetry between the left and right for $M\neq N$. If $M=N$ as well as the left-right symmetry is imposed, it must be spontaneously broken, along with the gauge symmetry breaking, in order to recover the standard model and the new physics besides. In other words, it all marks a theory that induces the observed left-right asymmetries. Further, our following discussions apply for all cases for $(M,N)=(2,3),\ (3,2),\ (3,3)$, and so forth, as well as for if the left-right symmetry is imposed or not, called non-minimal left-right gauge extensions.         

Denote the Cartan generators for $SU(M)_L$ and $SU(N)_R$ as $H_{iL}$~($i=0,1,2,...,M-2$) and $H_{jR}$~($j=0,1,2,...,N-2$), respectively. Further, we work in the basis of the generalized Gell-Mann matrices, i.e. $H_{0L,R}=T_{3L,R}$, $H_{1L,R}=T_{8L,R}$, and so forth (note that $H_k=T_{(k+2)^2-1}$). The electric charge operator is generally embedded as \be Q=H_{0L}+H_{0R}+\fr 1 2 (B-L),\ee where 
\be \fr 1 2 (B-L)\equiv \sum^{\mathrm{Max}\{M-2,N-2\}}_{k=1} \beta_k (H_{k L}+H_{k R})+X.\ee Here, note that either $H_{kL}=0$ for $k>M-2$ or $H_{kR}=0$ for $k>N-2$.    

The fermion content which is anomaly free is arranged as 
\bea  &&\psi_{aL}=
\left(
\begin{array}{c}
\nu_{aL}\\
e_{aL}\\
E^{q_1}_{aL}\\
\vdots \\
E^{q_{M-2}}_{aL}
\end{array}
\right)\sim \left(1,M,1,\fr{q_L-1}{M}\right),\crn
&&\psi_{aR}=
\left(
\begin{array}{c}
\nu_{aR}\\
e_{aR}\\
E^{q_1}_{aR}\\
\vdots \\
E^{q_{N-2}}_{aR}
\end{array}
\right)\sim \left(1,1,N,\fr{q_R-1}{N}\right),\eea
\bea
&& Q_{3L}=
\left(
\begin{array}{c}
u_{3L}\\
d_{3L}\\
J^{q_1+2/3}_{3L}\\
\vdots \\
J^{q_{M-2}+2/3}_{3L}
\end{array}
\right)\sim\left(3,M,1,\fr{q_L-1}{M}+\fr 2 3 \right),\crn
&& Q_{3R}=
\left(
\begin{array}{c}
u_{3R}\\
d_{3R}\\
J^{q_1+2/3}_{3R}\\
\vdots \\
J^{q_{N-2}+2/3}_{3R}
\end{array}
\right)\sim\left(3,1,N,\fr{q_R-1}{N}+\fr 2 3 \right),\eea
\bea 
&& Q_{\al L}=
\left(
\begin{array}{c}
d_{\al L}\\
-u_{\al L}\\
J^{-q_1-1/3}_{\al L}\\
\vdots \\
J^{-q_{M-2}-1/3}_{\al L}
\end{array}
\right)\sim\left(3,M^*,1,\fr{1-q_L}{M}-\fr 1 3 \right),\crn
&& Q_{\al R}=
\left(
\begin{array}{c}
d_{\al R}\\
-u_{\al R}\\
J^{-q_1-1/3}_{\al R}\\
\vdots \\
J^{-q_{N-2}-1/3}_{\al R}
\end{array}
\right)\sim\left(3,1,N^*,\fr{1-q_R}{N}-\fr 1 3 \right).
\eea The generation indices are $a=1,2,3$ and $\al=1,2$. The electric charge parameters satisfy $q_L=q_1+q_2+\cdots + q_{M-2}$ and $q_R=q_{1}+q_2+\cdots + q_{N-2}$. If $M=N$, $q_L=q_R$ and the fermion spectrum as given is completed. If $M\neq N$, $q_L\neq q_R$ and a number of chiral fermion partners have not been listed, which transform trivially under the left-right gauge groups (as concretely introduced in the following cases). The coefficients, $\beta_k$, can be determined through the parameters, $q_k$, as  
\be \beta_{k}=\sqrt{\fr{k}{k+2}}\beta_{k-1} +\sqrt{\fr{2(k+1)}{k+2}}(q_{k-1}-q_k),\ee where the initial conditions are $\beta_0=1$ and $q_0=-1$. We particularly provide $\beta_1=-(1+2q_1)/\sqrt{3}$, $\beta_2=-(1-q_1+3q_2)/\sqrt{6}$, and $\beta_3=-(1-q_1-q_2+4q_3)/\sqrt{10}$, which will appropriately be used for the realistic versions as shown below.     

Suppose that the nontrivial fermion representations under the left-right gauge groups are only defining representations or the complex conjugates of defining representations, respectively, and that they are enlarged from those of the standard model. The $SU(M)_L$ (or $SU(N)_R$) anomaly cancelation requires the number of fermion $M$-plets (or $N$-plets) to be equal to that of fermion anti-$M$-plets (or anti-$N$-plets), since a representation and its conjugate have opposite anomalies. Hence, the number of fermion generations must be a multiple of 3---fundamental color number. The QCD asymptotic freedom demands that the number of fermion generations is smaller than or equal to $33/(2\times \mathrm{Max}\{M,N\})=5,4,3$ for $\mathrm{Max}\{M,N\}=3,4,5$. Thus, the generation number is just three, as observed, and $M,N\leq 5$. That property for the standard model and the minimal left-right model disappears since the $SU(2)$ anomaly always vanishes for every fermion representation. Therefore, this approach provides a partial solution of the fermion flavor questions.          

The gauge symmetry breaking and mass generation are done by several scalar multiplets. The first one is 
\bea \phi =
\left(\begin{array}{ccccc}
\phi^0_{11} & \phi^+_{12} & \phi^{-q_1}_{13} & \cdots & \phi^{-q_{N-2}}_{1,N}\\
\phi^-_{21} & \phi^0_{22} & \phi^{-1-q_1}_{23} & \cdots & \phi^{-1-q_{N-2}}_{2,N}\\
\phi^{q_1}_{31} & \phi^{1+q_1}_{32} & \phi^{0}_{33} & \cdots & \phi^{q_{1}-q_{N-2}}_{3,N}\\
\vdots & \vdots & \vdots & \ddots & \vdots  \\
\phi^{q_{M-2}}_{M,1} & \phi^{1+q_{M-2}}_{M,2} & \phi^{q_{M-2}-q_1}_{M,3} & \cdots & \phi^{q_{M-2}-q_{N-2}}_{M,N}\\
\end{array}
\right)\sim \left(1,M, N^*, \fr{q_L-1}{M}-\fr{q_R-1}{N}\right),\eea which couples to $\bar{\psi}_{aL} \psi_{bR}$, $\bar{Q}_{3L}Q_{3R}$, and $\bar{Q}_{\al L}Q_{\beta R}$. The second one is 
\bea \Delta_R 
&=& \left(
\begin{array}{ccccc}
\Delta^{0}_{11} & \fr{1}{\sqrt{2}}\Delta^-_{12} & \fr{1}{\sqrt{2}} \Delta^{q_1}_{13} & \cdots & \fr{1}{\sqrt{2}} \Delta^{q_{N-2}}_{1,N}\\
\fr{1}{\sqrt{2}} \Delta^{-}_{12} & \Delta^{--}_{22} & \fr{1}{\sqrt{2}} \Delta^{-1+q_1}_{23} & \cdots & \fr{1}{\sqrt{2}} \Delta^{-1+q_{N-2}}_{2,N}\\
\fr{1}{\sqrt{2}} \Delta^{q_1}_{13} & \fr{1}{\sqrt{2}} \Delta^{q_1-1}_{23} & \Delta^{2q_1}_{33} & \cdots & \fr{1}{\sqrt{2}} \Delta^{q_1+q_{N-2}}_{3,N}\\
\vdots & \vdots & \vdots & \ddots & \vdots \\
\fr{1}{\sqrt{2}} \Delta^{q_{N-2}}_{1,N} & \fr{1}{\sqrt{2}} \Delta^{q_{N-2}-1}_{2,N} & \fr{1}{\sqrt{2}} \Delta^{q_{N-2}+q_1}_{3,N} & \cdots & \Delta^{2q_{N-2}}_{N,N}\\
\end{array}
\right)_R\crn
&\sim& \left(1,1,\fr{N(N+1)}{2},\fr{2(q_R-1)}{N}\right),\eea which couples to $\bar{\psi}^c_{aR}\psi_{bR}$. The multiplets, $\phi$ and $\Delta_R$, are minimally required if $M=N$.   

If $M>N$, the third one includes $M-N$ scalar multiplets, $\chi_p$ ($p=1,2,...,M-N$), where \bea \chi_p=\left(\begin{array}{c}
\chi^{-q_{N+p-2}}_1\\
\chi^{-1-q_{N+p-2}}_2\\
\chi^{q_1-q_{N+p-2}}_3\\
\vdots\\
\chi^0_{N+p}\\
\vdots\\
\chi^{q_{M-2}-q_{N+p-2}}_{M}
\end{array}
\right)\sim \left(1,M,1,\fr{q_L-1}{M}-q_{N-2+p}\right),\eea which couples to $\bar{\psi}_{aL} E^{q_{N+p-2}}_{bR}$, $\bar{Q}_{3L} J^{q_{N+p-2}+2/3}_{3R}$, and $\bar{Q}_{\al L} J^{-q_{N+p-2}-1/3}_{\beta R}$. Here, the second factors are the singlet fermion partners as mentioned. 

Otherwise, if $N>M$, the third one includes $N-M$ scalar multiplets instead, also denoted by 
\bea \chi_p  =  \left(\begin{array}{c}
\chi^{-q_{M+p-2}}_1\\
\chi^{-1-q_{M+p-2}}_2\\
\chi^{q_1-q_{M+p-2}}_3\\
\vdots\\
\chi^0_{M+p}\\
\vdots\\
\chi^{q_{N-2}-q_{M+p-2}}_{N}
\end{array}
\right) \sim \left(1,1,N,\fr{q_R-1}{N}-q_{M+p-2}\right),\eea which couples to $\bar{E}^{-q_{M+p-2}}_{aL}\psi_{bR}$, $\bar{J}^{-q_{M+p-2}-2/3}_{3L}Q_{3R}$, and $\bar{J}^{q_{M+p-2}+1/3}_{\al L} Q_{\beta R}$, where the first factors are the singlet fermion partners as mentioned, and in this case $p=1,2,3,...,N-M$. 

We see that $(\phi,\chi)$ for $M>N$ or $(\phi,\chi^\dagger)$ for $M<N$ always form a squared matrix of $\mathrm{Max}\{M,N\}$ dimension, in which every diagonal element is electrically neutral as the $M=N$ case is. Additionally, there might be two extra scalar multiplets (i) $\Delta_L$ that couples to $\bar{\psi}^c_{aL}\psi_{bL}$, defined similarly to $\Delta_R$ by replacing $(R,N)\rightarrow (L,M)$ and (ii) $\eta$ that couples $Q_{3L}$ to $Q_{\al R}$ as well as $Q_{3 R}$ to $Q_{\al L}$,
\bea \eta=\left(
\begin{array}{ccccc}
\eta^+_{11} & \eta^0_{12} & \eta^{1+q_1}_{13} &\cdots &\eta^{1+q_{N-2}}_{1,N}\\
\eta^{0}_{21} & \eta^{-1}_{22} & \eta^{q_1}_{23} &\cdots &\eta^{q_{N-2}}_{2,N}\\
\eta^{q_1+1}_{31} & \eta^{q_1}_{32} & \eta^{2q_1+1}_{33} &\cdots &\eta^{q_1+q_{N-2}+1}_{3,N}\\
\vdots & \vdots & \vdots & \ddots & \vdots\\
\eta^{q_{M-2}+1}_{M,1} & \eta^{q_{M-2}}_{M,2} & \eta^{q_{M-2}+q_1+1}_{M,3} &\cdots &\eta^{q_{M-2}+q_{N-2}+1}_{M,N}
\end{array}
\right)\sim \left(1,M,N, \fr{q_L-1}{M}+\fr{q_R-1}{N}+1 \right).\eea
Exceptionally, if either $M=2$ or $N=2$, we do not necessarily introduce $\eta$ because $\phi$ can play its role instead. Note that $\phi,\chi_p,\eta$ provides Dirac masses for fermions, while $\Delta_{L,R}$ provide Majorana masses for neutrinos. It is also shown that the gauge bosons can gain appropriate masses. The gauge symmetry of the model, $SU(3)_C\otimes SU(M)_L\otimes SU(N)_R\otimes U(1)_X$, is broken down to $SU(3)_C\otimes U(1)_Q\otimes P$, where $P$ is a discrete gauge symmetry.

Indeed, note that $\phi,\chi_p,\eta$ break the gauge symmetry down to $SU(3)_C\otimes U(1)_Q\otimes U(1)_{B-L}$, which leave $B-L$ conserved. This $B-L$ charge is only broken by $\Delta_{L,R}$. A transformation of $U(1)_{B-L}$ takes the form, $U(\omega)=e^{i\om (B-L)}$. $P$ is those $U(\om)$'s that conserve the vacuum, \bea \langle \Delta_{L,R}\rangle =\fr{1}{\sqrt{2}} \left(\begin{array}{cc}  v_{L,R} & \cdots \\ \vdots & \ddots \end{array}\right).\eea Since $[B-L](\Delta_{11})=-2$, we get $e^{-2i\om}=1$, which yields $\om = m\pi$ for $m=0,\pm1,\pm2,\cdots$, and thus $P=e^{im\pi(B-L)}=(-1)^{m(B-L)}$. Considering $m=3$ as well as multiplying the result by the spin parity, $(-1)^{2s}$, we obtain a conserved operator, $P=(-1)^{3(B-L)+2s}$, the so-called W-parity (since the wrong $B-L$ particles shown hereafter transform nontrivially under it).   

From the electric charge operator, the $B-L$ charges of the new fields that have $T_{3L,R}=0$ equal two times their electric charges, respectively, which is a new observation of this work. It is valid for every new fermion $(E,J)$ as well as numerous scalar and gauge boson fields. The $B-L$ charge and W-parity for all the fields are listed in Table \ref{tab1}. Here, we have denoted $P^\pm_k=(-1)^{\pm(6q_k+1)}$, which is nontrivial for $q_k\neq (2m-1)/6$, with $m=0,\pm1,\pm2,\cdots$ The particles that have nontrivial W-parity are named as W-particles since they possess wrong $B-L$ charges aforementioned, unlike the standard model particles. The particles that have $P=1$, which include the standard model particles and others, are called as normal particles. Thus, the normal particles also include those that have a $B-L$ charge differing from the usual one (defined by the standard model) by even unit, such as $\Delta_{11,12,22}$. It is noteworthy that if the electric charge parameters get values like ordinary electric charges, i.e. $q_k=m/3=0,\pm 1/3,\pm 2/3,\pm1,\cdots$, we have $P^\pm_k=-1$. In this case, W-parity behaves as R-parity, and the W-particles are just W-odd.     
\begin{table}[h]
\begin{tabular}{|l|cc|l|cc|}
\hline 
Field & $B-L$ & $P$ & Field & $B-L$ & $P$\\
\hline
$\nu_a$ & $-1$ & 1  &  $\Delta^{q_k+q_l}_{2+k,2+l}$ & $2(q_k+q_l)$ & $P^+_k P^+_l$  \\ 
$e_a$ & $-1$ & 1 &  $\chi^{-q_{N+p-2}}_1$ & $-1-2q_{N+p-2}$ & $P^-_{N+p-2}$  \\
$E^{q_k}_a$ & $2 q_k$ & $P^+_k$ &   $\chi^{-1-q_{N+p-2}}_2$ & $-1-2q_{N+p-2}$ & $P^-_{N+p-2}$   \\
$u_a$ & $1/3$ & 1 &  $\chi^{q_{k}-q_{N+p-2}}_{k+2}$ & $2(q_k-q_{N+p-2})$ & $P^+_k P^-_{N+p-2}$  \\
$d_a$ & $1/3$ & 1 & $\eta^+_{11}$ & 0 & 1 \\
$J^{q_k+2/3}_3$ & $2(q_k+2/3)$ & $P^+_k$ &  $\eta^0_{12}$ & 0 & 1  \\
$J^{-q_k-1/3}_\al $ & $-2(q_k+1/3)$ & $P^-_k$  & $\eta^0_{21}$ & 0 & 1 \\
$\phi^0_{11}$ & 0 & 1 & $\eta^-_{22}$ & 0 & 1 \\
$\phi^+_{12}$ & 0 & 1 & $\eta^{1+q_l}_{1,l+2}$ & $1+2q_l$& $P^+_l$ \\
$\phi^-_{21}$ & 0 & 1 & $\eta^{q_l}_{2,l+2}$ & $1+2q_l$ & $P^+_l$ \\
$\phi^0_{22}$ & 0 & 1 &  $\eta^{1+q_k}_{k+2,1}$ & $1+2q_k$ & $P^+_k$ \\
$\phi^{-q_l}_{1,2+l}$  & $-1-2q_l$ & $P^-_l$ &  $\eta^{q_k}_{k+2,2}$ & $1+2q_k$ & $P^+_k$ \\
$\phi^{-1-q_l}_{2,2+l}$ & $-1-2q_l$ & $P^-_l$  & $\eta^{q_k+q_l+1}_{k+2,l+2}$ & $2(q_k+q_l+1)$ & $P^+_k P^+_l$ \\
$\phi^{q_k}_{2+k,1}$ & $1+2q_k$ & $P^+_k$ & $\mathrm{gluon}$ & 0 & 1  \\
$\phi^{1+q_k}_{2+k,2}$ & $1+2q_k$ & $P^+_k$ & $B$ & 0 & 1 \\
$\phi^{q_k-q_l}_{2+k,2+l}$ & $2(q_k-q_l)$ & $P^+_k P^-_l$ & $A_{iL}$ & 0 & 1   \\
$\Delta^0_{11}$ & $-2$ & 1 & $A_{jR}$ & 0 & 1  \\
$\Delta^-_{12}$ & $-2$ & 1 & $W^{\pm}_{12}$ & 0 & 1 \\
$\Delta^{--}_{22}$ & $-2$ & 1 & $W^{-q_l}_{1,l+2}$ & $-1-2q_l$ & $P^-_l$ \\
$\Delta^{q_l}_{1,2+l}$ & $-1+2q_l$ & $P^+_l$ & $W^{-1-q_l}_{2,l+2}$ & $-1-2q_l$ & $P^-_l$ \\
$\Delta^{-1+q_l}_{2,2+l}$ & $-1+2q_l$ & $P^+_l$ & $W^{q_k-q_l}_{k+2,l+2}$ & $2(q_k-q_l)$ & $P^+_k P^-_l$ \\ 
\hline
\end{tabular}
\caption{\label{tab1} $B-L$ and W-parity for the model fields, where $k,l$ appropriately run from 1 to $M-2$ or $N-2$. The $\chi$ fields have been provided for $M>N$, while for $N>M$ those can be achieved by interchanging $M\leftrightarrow N$. We have also denoted $B$ as $U(1)_X$ gauge boson, $W_{L}$ ($W_R$) and $A_{L}$ ($A_{R}$) as non-Hermitian and neutral $SU(M)_L$ ($SU(N)_R$) gauge bosons, respectively. In the table, almost the $L,R$ notations for the relevant fields were omitted for brevity.}
\end{table}

The lightest W-particle (LWP) is stabilized due to W-parity conservation and kinematically suppressed. Further, there is no single W-field appearing in interactions since W-parity is conserved. If an interaction has $r_k$ of $P^+_k$ fields and $s_l$ of $P^-_l$ fields for some values of $k$ and $l$, where $r_k$ and $s_l$ are integer. The W-parity conservation implies $\sum_k r_k (6q_k+1)-\sum_l s_l (6q_l+1)=2m$ for some $m$ integer. This equation is valid for any $q_{k,l}$ parameter, which follows $k=l$ and $r_k=s_l$. Thus, $P^+_k$ and $P^-_k$ always appear in pairs in such interactions. If there are interactions that include only $r_{k l}$ of $P^+_{k}P^+_{l}$ fields and $s_{k' l'}$ of $P^-_{k'}P^-_{l'}$ fields, it deduces $\sum_{k,l}r_{k l}(6q_{k}+6q_{l})-\sum_{k',l'}s_{k' l'}(6q_{k'}+6q_{l'})=2m$, which is satisfied if $k=k'$, $l=l'$, and $r_{kl}=s_{k' l'}$. $P^+P^+$ and $P^-P^-$ always appear in pairs in such interactions. If an interaction contains only $t_{k l}$ of $P^+_{k}P^-_{l}$ fields, we have $\sum_{kl} t_{kl}(6q_{k}-6q_{l})=2m$, which implies $t_{kl}=t_{lk}$. Thus, $P^+P^-$ and its conjugate always appear in pairs in those interactions. We also see that the W-fields that have W-parity as $P^+P^-$, $P^+P^+$, or $P^-P^-$ often enter the self-interactions of three W-particles, where the last two have W-parity of $P^+$ and/or $P^-$.

If the LWP is electrically neutral, it may be a dark matter candidate. The models that provide dark matter candidates are listed in Table \ref{tab2}. We see that in the $q_k=0$ model, dark matter may be a neutral fermion, a neutral non-Hermitian gauge boson, or some neutral scalar. However, in the $q_k=-1$ model, dark matter include only a neutral non-Hermitian gauge boson or some neutral scalar. Whilst, the $q_k=1$ model yields dark matter particle uniquely as a neutral scalar.    
\begin{table}[h]
\begin{tabular}{|l|c|}
\hline
Model & Candidates\\
\hline
$q_k=0$ & $(E_a)_{k+2}$, $W_{1,k+2}$, $\phi_{1,k+2}$, $\phi_{k+2,1}$, $\Delta_{1,k+2}$, $\eta_{2,k+2}$, $\eta_{k+2,2}$, possibly $\chi_1$ \\
$q_k=-1$ & $W_{2,2+k}$, $\phi_{2,k+2}$, $\phi_{k+2,2}$, $\eta_{1,k+2}$, $\eta_{k+2,1}$, possibly $\chi_2$\\
$q_k=1$ & $\Delta_{2,k+2}$\\
\hline
\end{tabular}
\caption{\label{tab2} The dark matter models and corresponding candidates obtained when $q_{k}$ gets appropriate values for some $k$ values, respectively. The $\chi$ candidates are viable when either $p=k+2-N$ for $M>N$ or $p=k+2-M$ for $N>M$.} 
\end{table}                                       

{\it Realistic realizations}: They include the models that correspond to $(M,N)=(2,3)$, $(3,2)$, $(3,3)$,$\cdots$, or $(5,5)$ in order to contain W-particles as well as satisfying QCD asymptotic freedom. The one-parameter models $(q_1)$ coupled to $(M,N)=(2,3),\ (3,2)$, and $(3,3)$ were firstly introduced in \cite{hd} to solve the diphoton anomaly, which was an old story. Additionally, the implication for dark matter was firstly recognized in a particular setup for $M=2$ and $N=3$ \cite{dh1}. The two-parameter models $(q_1,q_2)$ are associated with $(M,N)=(2,4),\ (3,4),\ (4,2),\ (4,3),\ (4,4)$. We have seven models for the three parameters ($q_1,q_2,q_3$), where $M$ or $N$ equals 5.

Our proposal can naturally provide the small neutrino masses via seesaw mechanisms. It also supplies the tree-level flavor-changing neutral currents due to the non-universal couplings of the new neutral gauge bosons and scalars with ordinary quarks, which potentially explain the B physics anomalies (see \cite{dh1} and references therein). A further analysis yields that the normal particles have $Q$ and $B-L$ as quantized as usual values, while those charges of W-particles arbitrarily depend on the $q_k$ parameters (see \cite{d} for a similar treatment).                   

{\it Conclusion}: It has been shown that there is a bulk of wrong $B-L$ particles, which have $B-L$ related to the corresponding electric charge (most yield that $B-L$ equals two times the electric charge), responsible for dark matter. The candidate is the LWP which may include a neutral fermion, scalar, or vector, and is stabilized by W-parity.  

\section*{Acknowledgments}

This research is funded by Vietnam National Foundation for Science and Technology Development (NAFOSTED) under grant number 103.01-2016.77.

 \end{document}